# Reaction front occurrence on imperfection profiles during oxygen chemical diffusion in oxides. I. Thermodynamic background

M. Sinder*, Z. Burshtein, and J. Pelleg

*Materials Engineering Department, Ben-Gurion University of the Negev, P.O. Box 653, Beer-Sheva 84105, Israel*

Abstract

We present a theoretical study of the impact of oxygen diffusion in oxide crystals on metal dopants ionic state and the conduction type under thermal equilibrium. Oxygen vacancy formation acting as a shallow, double-electronic donor is assumed to result from the crystal exposure to a low ambient oxygen pressure. It is shown, that critical transitions from n- to a p-type at an oxygen partial pressure $P_i$, and in ionization state of the metal dopant at an oxygen partial pressure $P_M$, are usually not simultaneous, and depend on the different reaction constants. Experimental study of the different species concentrations at thermodynamic equilibrium as functions of pressure and temperature should allow assessment of various reversible reaction constants controlling the process. In the Part II companion paper, the kinetic (diffusion) characteristics are considered in detail.

*Corresponding author. email address: micha@bgu.ac.il

# I. INTRODUCTION

Oxide crystals are commonly used as hosts for different metal ions, serving various optical applications such as laser gain materials, and light saturable absorbers [1-3]. At elevated temperatures, the crystal may loose lattice oxygen under an ambient reducing atmosphere, thus creating oxide vacancies and free electrons. Oxygen diffusion in oxide crystals is of interest in relation to color centers [2], as well as valence transformations of embedded metal ions [3]. These processes belong to the general subject of chemical diffusion in solids, which is of considerable theoretical and practical importance [4-9].

In our present work we study the valence transformations of metal ions embedded in oxide crystals induced by slow changes in the ambient partial oxygen pressure. Critical transitions from n- to a p-type at an oxygen partial pressure $P_i$, and in ionization state of the metal dopant at an oxygen partial pressure $P_M$, are usually not simultaneous, and depend on the different reaction constants

We consider the issue in two companion papers; the present one, Part I, provides a thermodynamic analysis, namely the concentration of the various species involved in the physical and chemical process after achievement of thermodynamic equilibrium. From an experimental point of view it relates to very small rates of the ambient oxygen pressure changes. The material shown is in fact wholly contained in reference [8], and is provided here separately for completeness of presentation and ease of comprehension of our next, Part II companion paper. The said Part II companion paper, addresses the kinetic aspects, namely the temporal changes and spatial profiles developing during the oxygen vacancy diffusion upon sudden gross changes in the ambient oxygen pressure.

## II. MODEL

The oxygen vacancies involved are assumed to be shallow, each contributing two electrons to the lattice conduction band. The oxygen surface evaporation is described by the reaction

$$O^{2-}(\text{solid}) \leftrightarrow \tfrac{1}{2}O_2(\text{gas}) + V_O + 2e^-, \qquad (1)$$

where $O^{2-}$ is a lattice oxygen atom, $e^-$ is a free electron, and $V_O$ is a doubly ionized oxygen vacancy. The free electron-hole reaction is given by

$$e^- + h^+ \leftrightarrow 0, \qquad (2)$$

where $h^+$ is a free hole. We assume that the metallic dopant ion $M^{p+}$ is a deep acceptor; namely, it may capture a single free electron and change into an $M^{(p-1)+}$ ion:

$$M^{p+} + e^- \leftrightarrow M^{(p-1)+}. \qquad (3)$$

Fig. 1 illustrates the energy level scheme of our model.

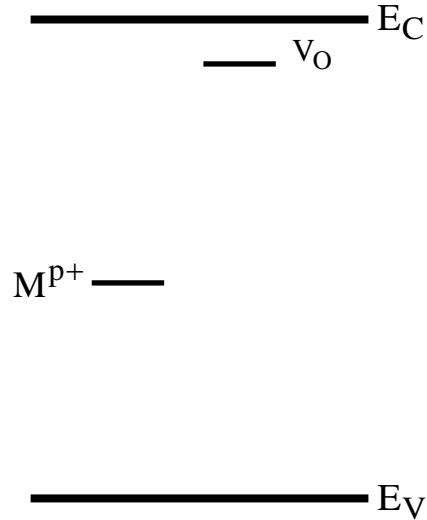

FIG. 1 : Energy scheme for an oxide crystal, deep $M^{p+}$- dopant, and a shallow oxygen vacancy $V_O$ considered in our analysis. $E_V$ and $E_C$ are the top valence and bottom conduction band energies, respectively.

## III. THERMODYNAMIC ANALYSIS

Applying the law of mass action to Eqs. (1)-(3) yields

$$\frac{P^{1/2}[V_O][e^-]^2}{[O^{2-}]} = K_1 \quad ; \tag{4}$$

$$[e^-][h^+] = K_2 \quad ; \tag{5}$$

$$\frac{[M^{p+}][e^-]}{[M^{(p-1)+}]} = K_3 \quad , \tag{6}$$

where $K_1$, $K_2$, and $K_3$ are the corresponding temperature-dependent reaction constants. In addition, the electro-neutrality condition is

$$[h^+] + 2[V_O] = [e^-] + [M^{(p-1)+}] \quad , \tag{7}$$

and the M-dopant conservation condition

$$[M^{p+}] + [M^{(p-1)+}] = [M] \quad . \tag{8}$$

Equations (4)-(8) establish a set of 5 independent equations for the 5 unknown variables $[M^{p+}]$, $[M^{(p-1)+}]$, $[h^+]$, $[e^-]$, and $[V_O]$. A reduced form of these variables is

$$\mu \equiv [M^{p+}]/[M], \quad \bar{\mu} \equiv [M^{(p-1)+}]/[M], \quad h \equiv [h^+]/[M], \quad \varepsilon \equiv [e^-]/[M] \text{ and } v \equiv [V_O]/[M].$$

A parametric solution for the reduced variables as functions of the ambient partial oxygen pressure $P$, using $\varepsilon$ as a free parameter, is

$$\mu = \frac{1}{1 + \varepsilon/\tilde{K}_3}; \quad \bar{\mu} = 1 - \mu; \quad h = \tilde{K}_2/\varepsilon; \quad v = \frac{1}{2}(\varepsilon - h + \bar{\mu}); \quad \tilde{P} = \frac{1}{v^2 \varepsilon^4} \quad , \tag{9}$$

where $\tilde{P} \equiv P \cdot \frac{[M]^6}{K_1^2 [O^{2-}]^2}$, $\tilde{K}_2 \equiv K_2/[M]^2$ and $\tilde{K}_3 \equiv K_3/[M]$, are respectively the ambient oxygen pressure and the reaction constants in dimensionless forms.

Based on Eq. (9), the different species concentrations as functions of the partial ambient oxygen pressure are presented in Fig. 2 for different values of $K_2$ and $K_3$. In Fig. 2(a), $K_3^2$ is taken very small compared to $K_2$, a condition automatically satisfied when $M^{p+}$ ions reside below mid-gap (Fig. 1); In contrast, in Fig. 2(b), $K_3^2$ is taken very large compared to $K_2$, a condition satisfied when $M^{p+}$ ions reside above mid-gap.

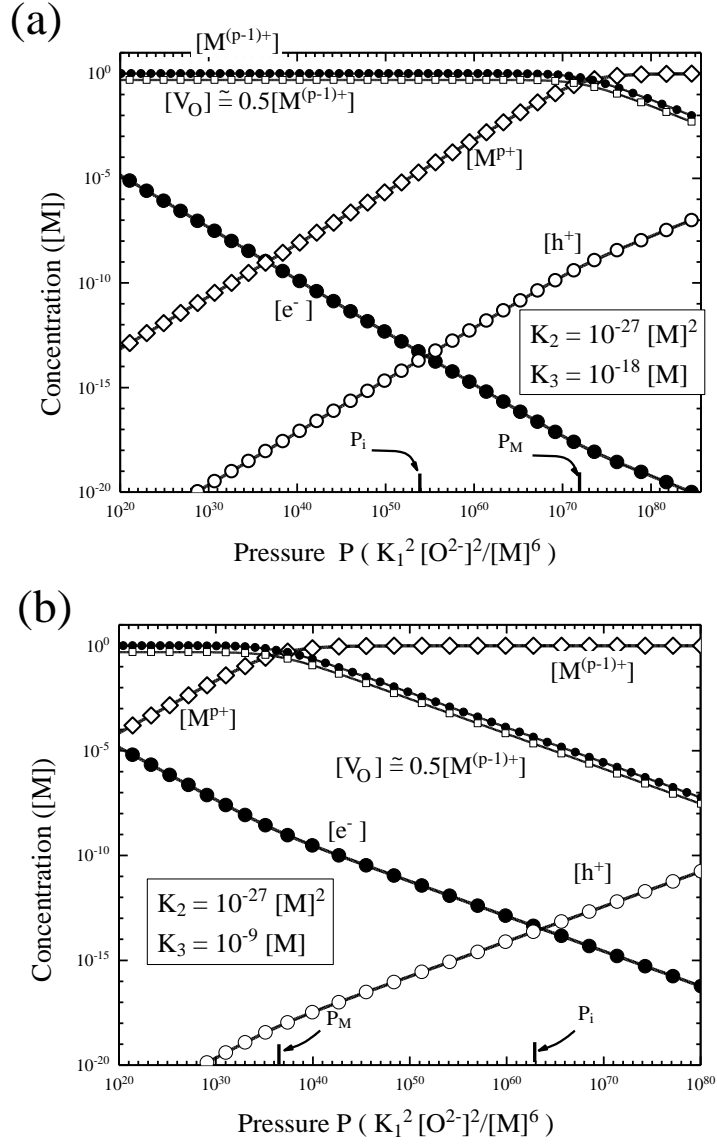

FIG. 2: Concentrations of dopant metal ion at its two ionization states $M^{p+}$ and $M^{(p-1)+}$, the doubly ionized vacancy $V_O$, electrons $e^-$, and holes $h^+$ for an oxide crystal host of 8 eV energy gap as functions of the partial ambient atmosphere oxygen pressure. (a) The $M^{p+}$ ion state located 1.0 eV below mid-gap; (b) The $M^{p+}$ state located 1.0 eV above mid-gap. The reaction constants relevant to each case are inset.

The specific parameter values used in Fig. 2(a), are $K_2 = 10^{-27} \, [M]^2$ and $K_3 = 10^{-18} \, [M]$; both correspond to the values $K_2 = 10^9 \, cm^{-6}$ and $K_3 = 1 \, cm^{-3}$ obtained for an assumed case where $[M] \approx 10^{18} \, cm^{-3}$, $N_C = N_V = 10^{22} \, cm^{-3}$, $E_g = 8 \, eV$, $k_B T = 0.1 \, eV$, and the $M^{p+}$ ion energy state located $\cong 1 \, eV$ <u>below mid-gap</u>, where $N_C$ and $N_V$ are the conduction-, and valence-band densities of states, respectively, and $E_g$ is the oxide matrix band-gap. As the oxygen partial pressure lowers, the dopant ion $M^{p+}$ concentration reduces, since some transform into $M^{(p-1)+}$ ones. Following the electro-neutrality condition (Eq. (7)), the reduced ion concentration $[M^{(p-1)+}]$ equals $2[V_O]$. The crossing between the $[M^{p+}]$ and $[M^{(p-1)+}]$ curves defines a characteristic pressure $P_M$. For pressures lower than $P_M$, most ions are $M^{(p-1)+}$; for pressures higher than $P_M$, most ions are $M^{p+}$. In addition, a crossing between the $[h^+]$ and $[e^-]$ curves defines a characteristic pressure $P_i$, residing approximately 18 orders of magnitude below $P_M$. For pressures lower than $P_i$, the crystal is electrically n-type; above $P_i$ it is electrically p-type.

In Fig. 2(b) we assumed $K_2 = 10^{-27} \, [M]^2$ and $K_3 = 10^{-9} \, [M]$; both correspond to the values $K_2 = 10^9 \, cm^{-6}$ and $K_3 = 10^9 \, cm^{-3}$ obtained for an assumed case same as in Fig. 2(a), yet the $M^{p+}$ ion energy state is located $\cong 1 \, eV$ <u>above mid-gap</u>. The characteristics differ grossly from those of Fig. 2(a). Especially, the order between $P_i$ and $P_M$ is reversed, namely $P_i$ now resides approximately 27 orders of magnitude above $P_M$.

Experimental measurement of the sample physical parameters at the characteristic situations defining $P_i$ and $P_M$ should be very valuable in assessing the three reaction constants $K_1$, $K_2$ and $K_3$. In principle, however, only two of the above may be inferred from two-measured characteristic pressures, while a third one, say $K_2$, may be assessed theoretically. Some useful relations obtained from Eq. (9) are, $K_3 = [e^-]_M$ and $K_2 = [e^-]_i^2$, where $[e^-]_M$ and $[e^-]_i$ are the electron concentrations at the characteristic pressures $P_M$ and $P_i$, respectively; also $P_M = 16(K_1[O^{2-}][M]^{-1})^{+2} K_3^{-4}$ and

$$P_i = 4(K_1 K_3 [O^{2-}][M]^{-1} K_2^{-3/2})^2 \left(1 + \sqrt{K_2}/K_3\right)^2.$$

Tables 1 and 2 summarize the energy schemes, reaction parameters and characteristic pressures used for the different demonstrations in the present paper Part I, as well as its companion paper Part II.

TABLE 1: Summary of energy schemes, reduced reaction parameters, and characteristic pressures, used for demonstrations. All relate to an energy gap $E_g = 8\,eV$, lattice oxygen ion concentration $[O^{2-}] = 10^{22}\,cm^{-3}$, temperature $k_BT = 0.1\,eV$, and total metal ion concentration $[M] = 10^{18}\,cm^{-3}$. All characteristic pressures $P_M$ relate to an actual oxygen partial pressure of 1 atm. For definition of the various parameters – see text.

| $M^{p+}$ ion energy position | Energy scheme | $\tilde{K}_2$ | $\tilde{K}_3$ | $\tilde{P}_i$ | $\tilde{P}_M$ |
|---|---|---|---|---|---|
| 1 eV above mid-gap | 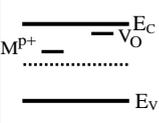 | $10^{-27}$ | $10^{-9}$ | $4 \times 10^{63}$ | $2 \times 10^{37}$ |
| 1 eV below mid-gap | 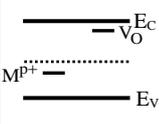 | $10^{-27}$ | $10^{-18}$ | $4 \times 10^{54}$ | $2 \times 10^{73}$ |

TABLE 2: Summary of energy schemes, reaction parameters, and characteristic pressures used for demonstrations. All relate to an energy gap $E_g = 8\,eV$, lattice oxygen ion concentration $[O^{2-}] = 10^{22}\,cm^{-3}$, temperature $k_BT = 0.1\,eV$, and total metal ion concentration $[M] = 10^{18}\,cm^{-3}$. All characteristic pressures $P_M$ relate to an actual oxygen partial pressure of 1 atm. For definition of the various parameters – see text.

| $M^{p+}$ ion energy position | Energy scheme | $K_1$ | $K_2$ | $K_3$ | $P_i$ | $P_M$ |
|---|---|---|---|---|---|---|
| 1 eV above mid-gap | 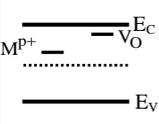 | $0.1\,g^{1/2}s^{-1}cm^{-13/2}$ | $10^9\,cm^{-6}$ | $10^9\,cm^{-3}$ | $2 \times 10^{26}$ atm. | 1.0 atm. |
| 1 eV below mid-gap | 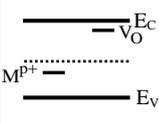 | $0.1\,g^{1/2}s^{-1}cm^{-13/2}$ | $10^9\,cm^{-6}$ | $1\,cm^{-3}$ | $2 \times 10^{-19}$ atm. | 1.0 atm. |

## V. CONCLUSIONS

In our present work we consider the oxygen vacancy to be always doubly ionized. An electron capture by the dopant metal ion involves interaction with the free electron-hole system. Oxidation and reduction processes of the metal ions are induced by slow changes in the ambient partial oxygen pressure. Experimental study of the different metal concentrations in the two ionic states at thermodynamic equilibrium as functions of pressure and temperature should allow assessment of various reaction constants controlling the process. Particularly, measurements of a sample physical parameters at the situations defining the critical partial oxygen pressures $P_i$ and $P_M$ would yield the three reaction constants $K_1$, $K_2$, and $K_3$. In principle, only two of the above may be inferred from the two-measured critical pressures, while a third one, say $K_2$, may be assessed theoretically. For example, the dopant metal concentrations at the two ionic states may be studied even after quenching to room temperature by electron paramagnetic resonance (EPR), and/or absorption or fluorescence spectroscopy.

**Figure captions**

FIG. 1 : Energy scheme for an oxide crystal, deep $M^{p+}$-dopant, and a shallow oxygen vacancy $V_O$ considered in our analysis. $E_V$ and $E_C$ are the top valence and bottom conduction band energies, respectively.

FIG. 2: Concentrations of dopant metal ion at its two ionization states $M^{p+}$ and $M^{(p-1)+}$, the doubly ionized vacancy $V_O$, and electrons $e^-$ and holes $h^+$ for an oxide crystal host of $8\,eV$ energy gap as functions of the partial ambient atmosphere oxygen pressure. (a) The $M^{p+}$ ion state located $1.0\,eV$ below mid-gap; (b) The $M^{p+}$ state located $1.0\,eV$ above mid-gap. The reaction constants relevant to each case are inset.